# About the significance of the driving current direction in ferromagnetic resonance experiments

Md. Majibul Haque Babu[1] and Maxim Tsoi[1,2,a)]

[1]*Department of Physics, University of Texas at Austin, Austin, TX 78712, USA*

[2]*Texas Materials Institute, University of Texas at Austin, Austin, TX 78712, USA*

## ABSTRACT

We present an experimental study of the effects of driving current direction on ferromagnetic resonance in NiFe foils. The rf driving current was applied to NiFe foils of different shapes. In rectangular samples with a close-to-uniform flow of the applied current along the long edge of the sample we find the resonance field to follow a simple '*cos*' dependence on the angle between the current and external dc magnetic field. We argue that this behavior cannot be explained by the in-plane demagnetizing field of the rectangular foil. In triangular samples where the current partially flows along all three sample edges we observed three independent '*cos*' features. The latter suggests individual contributions from different areas with different current directions. We were able to switch off one of these contributions by covering one edge of the triangular sample with a conducting overlayer and thereby effectively short-circuiting the corresponding current path. Our findings highlight the significance of driving current distributions in ferromagnetic resonance experiments.

## I. Introduction

Ferromagnetic resonance (FMR) [1] is a powerful experimental technique that probes the magnetization dynamics in various magnetic media, from bulk ferromagnets to nanoscale heterostructures [2-9]. The resonance occurs when an externally applied rf magnetic field drives the media's magnetization into precession at its natural frequency. The rf field is often produced by an applied rf current flowing through the media, thus, making the direction of this driving current an important parameter in FMR experiments. We have recently observed that FMR in a NiFe wire [10] depends on the angle between the driving current and external dc magnetic field. However, this result was inconclusive because a similar behavior could be associated with the demagnetizing field of the wire. Here we study FMR in NiFe thin foils with different shapes to address the question of demagnetizing field.

In our experiments FMR is driven by rf current applied to NiFe foils with rectangular and triangular shapes. The resonance occurs when the current's frequency matches the natural precession frequency of the foil's magnetization. The latter is defined by an external dc magnetic field applied in the sample's plane in different directions. We find that the resonance field increases significantly when the dc field is applied perpendicular to sample edges. The resonance field follows a simple '*cos*' dependence on the angle between the external dc field and orientation of the edge. An increase of the resonance field could in principle be associated with demagnetizing field, which sets in when the sample's magnetization is perpendicular to the edge. However, our analysis of demagnetizing factors in the simple rectangular geometry rules out such an explanation. Instead, we argue that the resonance field is controlled by the direction of current flow in the sample that includes the edges. We have attempted to effectively short-circuit the current flow along one of edges in our triangular sample by covering the edge with a conducting silver paste. This effectively switched off the FMR contribution from this edge, thus, highlighting the importance of the driving current distribution in FMR experiments.

## II. Methods

To observe FMR we used a 50 μm thick $Ni_{36}Fe_{64}$ foil (Goodfellow FE02-FL-000140). This is a very common ferromagnetic material, so we expect our results to generalize to many other materials as well. Here we focus on a sample with rectangular (2.33mm×1.24mm) shape. In our experimental setup [see Fig. 1(a)], the sample terminates a coaxial cable which is used to deliver rf current at frequency $f$ = 4.5-11 GHz from a microwave source (Agilent E8257D) to the sample. The reflected microwave power is then measured by a power sensor (Keysight U2002A) as a function of dc magnetic field $B$ applied in the sample plane. The reflected signal is directed to the sensor via a directional coupler (Mini-Circuits ZUDC20-183+). All measurements were performed at room temperature.

Figure 1(b) shows an example of such a conventional absorption measurement where the reflected power is a function of applied field $B$. The dips at $B = \pm 0.06$ T correspond to FMR at 7 GHz. Figure 1(c) shows that the FMR frequency follows the Kittel's [1] dispersion relation $(f/\gamma)^2 = [B + (N_z - N_x)\mu_0 M_s][B + (N_y - N_x)\mu_0 M_s]$, where $\gamma$ = 28 GHz/T is the gyromagnetic ratio, $\mu_0 M_s$ = 1.2 T the saturation magnetization, and ($N_x$, $N_y$, $N_z$) are demagnetizing factors which account for the effects of sample shape on the resonance frequency. The black curve in Fig. 1(c) shows a fit to the experimental data for factors (0.014, 0.006, 0.960).

In our setup the dc magnetic field can be rotated by 360 degrees in the sample plane. We can thus detect any variations in the resonance field as a function of its direction. For a circular thin-film sample without in-plane anisotropies one expects no variations in the resonance field for such a rotation – all in-plane directions are equivalent. For a rectangular sample, however, one can expect a change of the resonance field due to a variation in the demagnetizing field, e.g., when the applied dc field saturates the sample's magnetization perpendicular to its long edge, a larger demagnetizing field results in a larger resonance field. To estimate such a change, we calculated the demagnetizing factors for our sample (2.33 mm×1.24 mm×50 μm) by using the general expression (Eq. 1 in Ref. 11) for demagnetizing factors ($N_x$, $N_y$, $N_z$) of a rectangular ferromagnetic prism [11, 12] and found (0.028, 0.055, 0.917). The red and green curves in Fig. 1(c) show the dispersion relations for dc field applied parallel and perpendicular to the long edge of our rectangular sample. These estimations suggest that for different directions of the in-plane magnetic field the resonance field can vary by as much as 0.06 T. In what follows we present our angular dependent FMR measurements for the rectangular sample and compare our results with these expectations. We will further investigate the effects of sample shape on FMR frequency by performing the angular dependent FMR measurements in a triangular (2.33mm×1.52mm×1.4mm) sample [see Fig. 1(d)].

We will also compare our angular dependent results with a simple '*cos*' model proposed in [10]. The model highlights the significance of the driving rf current direction. It assumes that when the dc magnetic field is applied at an angle $\theta$ with respect to the rf current, which flows through a sample and drives FMR, only the perpendicular to current component of the field will contribute to FMR and the angular dependence of resonance field follows a simple relationship $B(\theta)=B(\theta=0°)/cos\theta$ [10], which we call the '*cos*' model.

## III. Results and Discussion

Figure 2(a) shows the angular dependence of FMR in the rectangular sample at $f=7$ GHz. The 2D gray-scale plot shows the FMR absorption spectra as a function of angle $\theta$, which designates the direction of the in-plane field *B*. The angle $\theta = 0°$ corresponds to the field directed along the long edge of rectangular sample [see Fig. 1(a)]. The blue horizontal line in Fig. 2(a) indicates the position of the absorption spectrum at $\theta = 0°$ shown in Fig. 1(b). A darker color in the gray-scale plot indicates a lower reflected power (higher absorption) and highlights variations in the resonance field as a function of $\theta$. The red curve is the '*cos*' model fit, which somewhat hides from view the experimental data at negative fields, but as the gray-scale plot is symmetric about $B = 0$ the '*cos*' pattern in experimental $B(\theta)$ data at positive fields is unobscured by the fit. Overall, the '*cos*' model fits the experimental data well, while some deviations may be associated with a non-uniform current flow in our sample, in particular around the contact electrodes where rf current enters/leaves the sample. The resonance field for $\theta = 0°$ is 0.06 T. When the direction of applied field approaches $\theta = 90°$, the resonance field increases by 0.14. This is significantly higher than the increase expected from the demagnetizing field in our rectangular sample. In contrast, this increase of the resonance field correlates well with the '*cos*' model. Figure 2(b) shows that the resonance field follows the '*cos*' model up to the highest fields (0.4 T) applied in our experiments. This is almost a factor of 7 higher than the expectations from demagnetizing fields. Note that this increase ('*cos*' pattern) occurs when the applied dc field is perpendicular to the sample's edge, i.e., perpendicular to the rf driving current [red arrow in Fig. 1(a)], that strongly supports the '*cos*' model.

Figure 3(a) shows the angular dependence of FMR in the triangular sample at $f=7$ GHz. The 2D gray-scale plot shows the FMR absorption spectra for different directions ($\theta$) of the in-plane field *B*. Here, $\theta = 0°$ corresponds to the field directed along the longest edge of triangular sample [see Fig. 1(d)]. Like in Fig. 2 a darker color highlights variations in FMR field as a function of $\theta$. For the triangular sample we could identify three different '*cos*' patterns (similar to the one in Fig. 2). Red, blue, and green curves in Fig. 3(a) are the '*cos*' model fits. The red '*cos*' pattern (increase of the FMR field) occurs when the applied dc field is perpendicular to the longer edge of the sample ($\theta = 90°$). The blue and green curves are shifted (vertically) with respect to the red curve by +30 and -35 degrees, respectively. Note [see Fig. 1(d)] that the two shifts correspond well to the orientations of the other two (shorter) sample edges (+34 and -38 degrees). This observation suggests that the three '*cos*' features (red, blue, green) may be associated with the three edges of our sample. Every time the applied magnetic field is perpendicular to one of the edges [red, blue and green arrows in Fig. 1(d)] the resonance field increases according to the '*cos*' model, just like in the case of the rectangular sample (Fig. 2). The red '*cos*' model fit in Fig. 3(a) highlights the resonance associated with the driving current flow along the longest edge of our triangular sample, while the other two (blue and green) '*cos*' features can be associated with currents along the other two edges of the sample. To verify this association between different '*cos*' patterns and the sample edges we have attempted to switch off a contribution to FMR from one of the edges as we show next.

We have used a conducting silver paste to effectively short circuits the current path along one of the edges. The dashed oval in Fig. 1(d) shows the location where the paste was applied to the sample as a conducting overlayer. Our intention was to prevent the rf current from flowing through the sample along this edge, but instead to be redirected into the silver overlayer. The rf current, which flows along this edge through the overlayer, will not contribute to FMR and we expect the corresponding '*cos*' pattern to vanish. Figure 3(b) shows the angular dependence of FMR in the triangular sample (at $f$=7 GHz) with the silver paste applied to one of the edges. This 2D gray-scale plot shows that one of the '*cos*' patterns (blue) of Fig. 3(a) has indeed vanished from the picture and does not appear in Fig. 3(b) [compare Figs. 3(a) and 3(b)]. This observation can be used as additional evidence of the important role played by the direction of driving current in FMR experiments.

## IV. Summary

We have experimentally investigated the angular dependence of FMR in $Ni_{36}Fe_{64}$ foils with different shapes. In a rectangular sample, we observed the resonance field to increase significantly when the dc magnetic field is applied perpendicular to the flow of the rf current driving FMR. This increase cannot be explained by the in-plane demagnetizing field but follows well the '*cos*' model, which highlights the significance of the driving current direction. In a triangular sample, we observed similar increases in FMR field for all three sample edges. The contribution from one of the sample edges could then be switched off by blocking the current flow along this edge. Our results highlight the importance of the relative orientation between the driving rf current and dc magnetic field in FMR experiments.

This work was supported in part by the University of Texas at Austin OVPR Special Research Grant.

# References

1. C. Kittel, *Introduction to Solid State Physics*, 8th ed, Wiley, Hoboken, NJ, 2005.
2. J. C. Sankey et al., *Phys. Rev. Lett.* **96**, 227601 (2006).
3. T. Staudacher and M. Tsoi, *J. Appl. Phys.* **109**, 07C912 (2011).
4. L. Liu et al., *Phys. Rev. Lett.* **106**, 036601 (2011).
5. M. Weiler et al., *Phys. Rev. Lett.* **106**, 117601 (2011).
6. D. Fang et al., *Nat. Nanotechnol.* **6**, 413 (2011).
7. C. Wang, H. Seinige, M. Tsoi, *J. Phys. D: Appl. Phys.* **46**, 285001 (2013).
8. H. Seinige, C. Wang, M. Tsoi, *Proc. SPIE* **8813**, 88131K (2013).
9. C. Wang, H. Seinige, M. Tsoi, *Low Temp. Phys.* **39**, 247 (2013).
10. Q. Gao and M. Tsoi, *J. Magn. Magn. Mater.* **580**, 170947 (2023).
11. A. Aharoni, *J. Appl. Phys.* **83**, 3432 (1998).
12. R. I. Joseph and E. Schlömann, *J. Appl. Phys.* **36**, 1579 (1965).

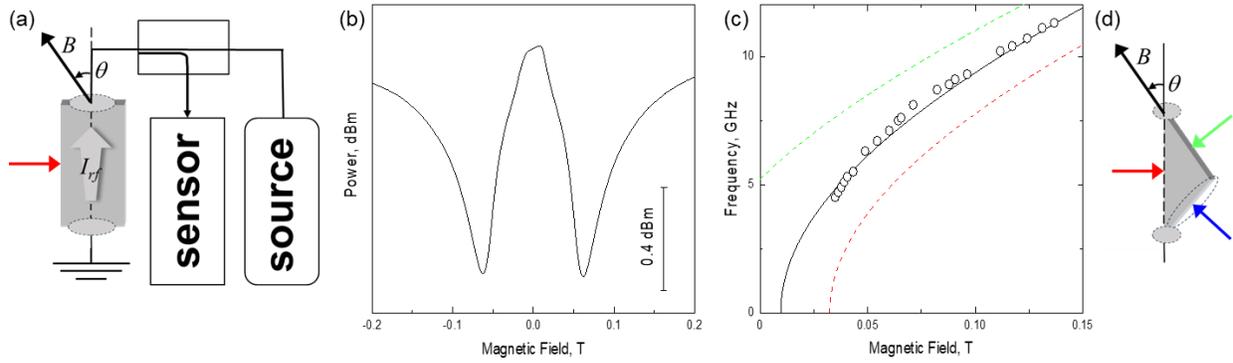

**Figure 1**. (a) Experimental setup. A rectangular thin-film sample terminates a coaxial cable used to deliver rf current from a microwave source to the sample. The reflected power is measured by a power sensor. Magnetic field *B* is applied in the sample plane at an angle $\theta$ with respect to the long edge of the sample. The rf current $I_{rf}$ flows in the sample plane between two solder contacts (gray ovals). (b) Absorption spectrum shows FMR around ±0.06 T at 7 GHz. (c) FMR dispersion (resonance field vs applied frequency). The experimental data (open circles) fitted by Kittel's relations (see text for details). (d) Triangular sample with in-plane dc field *B* applied in the sample plane at an angle $\theta$ with respect to the long edge of the sample. Gray ovals indicate contacts. Dashed oval indicates placement of conducting overlayer on one edge.

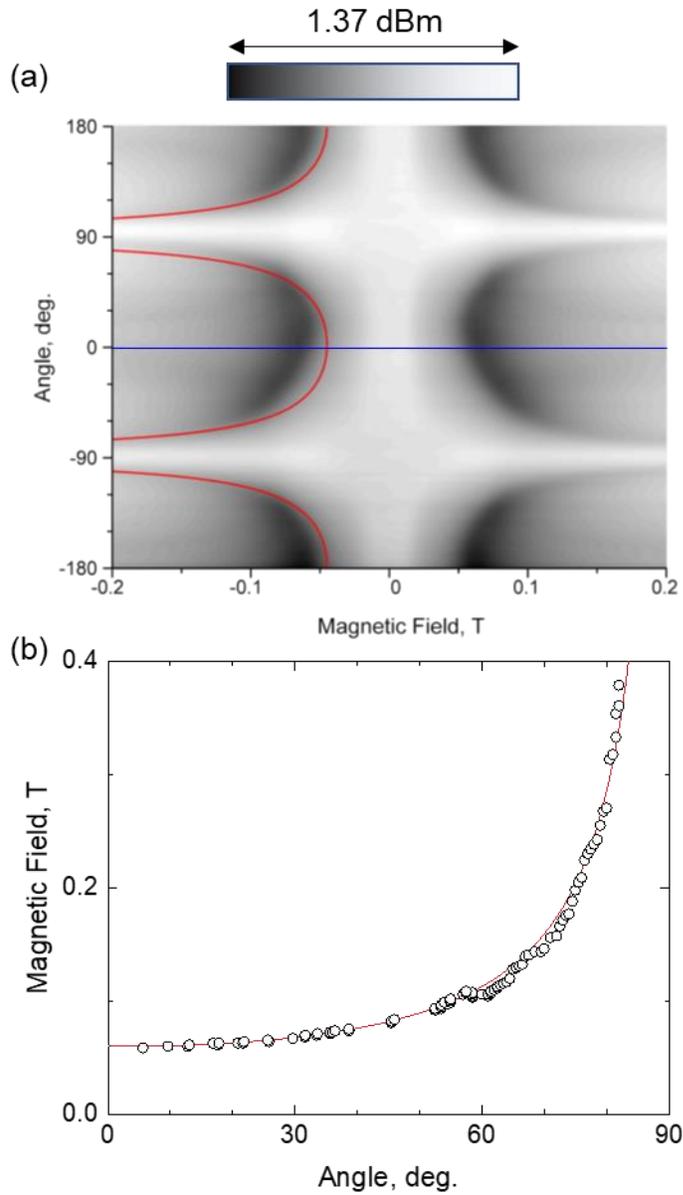

**Figure 2**. (a) 2D gray-scale plots show the FMR absorption spectra for rectangular sample as a function of the magnetic field angle $\theta$. The frequency of applied microwaves is 7 GHz. Darker color indicates lower power (higher absorption); the scale bar shows the black/white range. The blue horizontal line indicates the position of $\theta = 0°$ spectrum from Fig. 1(b). The red curve is the '*cos*' model fit. (b) Angular dependence of FMR ($B$ vs $\theta$) with the '*cos*' model fit (red curve).

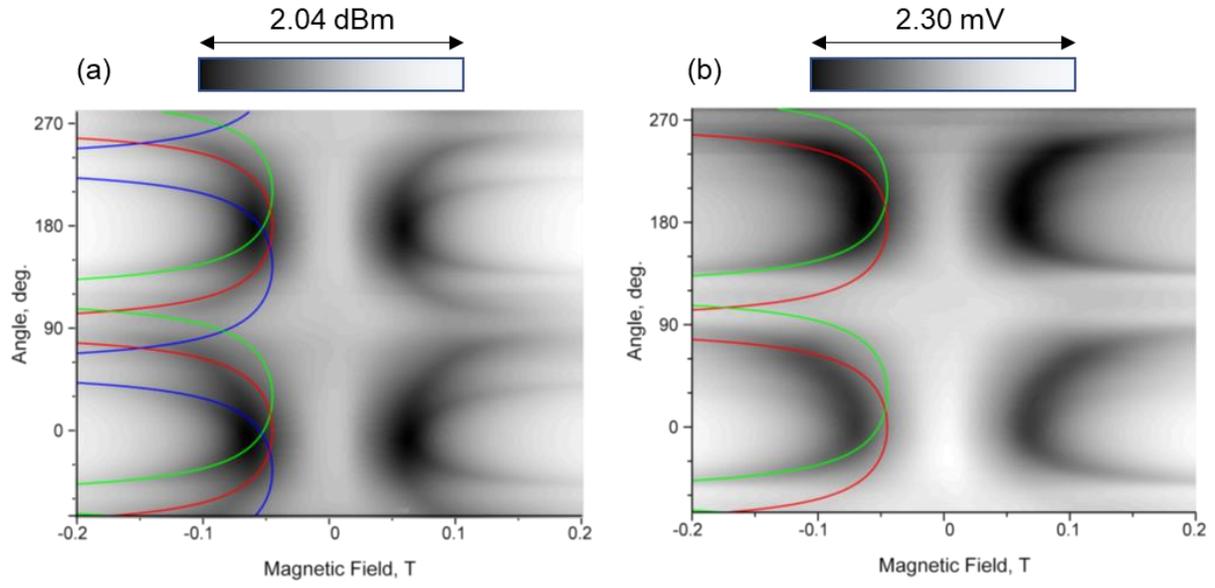

**Figure 3**. 2D gray-scale plots show the FMR absorption spectra as a function of the magnetic field angle $\theta$ for the triangular sample with (b) and without (a) a conducting overlayer on one of the sample edges. The frequency of applied microwaves is 7 GHz. Darker color indicates lower power (higher absorption); the scale bars show the black/white ranges. The red, blue, and green curves are the '*cos*' model fits. The blue feature vanishes in (b).